\documentclass[conference]{IEEEtran}
\usepackage{amsmath,amssymb,amsfonts}

\usepackage{xcolor}
\ifCLASSINFOpdf
\else
\fi
\usepackage{array}

\usepackage{graphicx}

\usepackage{amsmath}
\usepackage{amssymb}
\usepackage{booktabs}

\begin{document}
%




\title{CarnegiePLUG:\\
Prosumer-in-the-Loop simUlation Grid}


\author{
\IEEEauthorblockN{Juye Kim}
\IEEEauthorblockA{Electrical and Computer Engineering\\Carnegie Mellon University\\
juyek@andrew.cmu.edu}
\and
\IEEEauthorblockN{Carolyn Goodman}
\IEEEauthorblockA{College of Engineering\\
Carnegie Mellon University\\
cgoodman@andrew.cmu.edu }
\and
\IEEEauthorblockN{Javad Mohammadi}
\IEEEauthorblockA{Electrical and Computer Engineering\\Carnegie Mellon University\\
jmohamma@andrew.cmu.edu}
}


%


\maketitle

\begin{abstract}
This paper introduces Carnegie Mellon campus-wide Carnegie$\mathbb{PLUG}$ test-bed. This test-bed in a hardware-in-the-loop simulator that enables large-scale sensing, computation and actuation over a network of heterogeneous energy producers and consumers (prosuming) assets. Carnegie$\mathbb{PLUG}$ is a multi-agent framework that facilitates plug-and-play integration of a wide range of virtual and physical sensors and controllers through a multi-layer architecture. Virtual functionalities are accessed in software environments that interface with communication layer of our test-bed. This paper discusses Carnegie$\mathbb{PLUG}$ layers, provides implementation details and showcases potential applications of this test-bed through preliminary results.

\end{abstract}

\textit{\textbf{Keywords- Distributed Energy Resources, IoT-connected energy prosumers (producer and consumer)}}

%
\IEEEpeerreviewmaketitle

\section{Introduction}
\subsection{Motivation}
It is expected that the urban electric grid of the future smart cities would differ from the current system by the increased integration of Distributed Energy Resource (DER)s and communication and sensing technologies. This transition from the operational perspective means increased uncertainty and more actuation points and control decisions to make. A key question that needs to be answered is how the sensing and networking can be used efficiently to ensure a reliable and secure operation of the urban electric power networks despite the increased cyber vulnerabilities and challenges imposed by inherent intermittency of  DERs, e.g., rooftop solar PVs. 
Multi-agent and decentralized monitoring and control is key to achieve scalability.
We have developed a scalable sensing and decision-making test-bed that integrates coordination with ad-hoc networking and enables geographically spread energy producing and consuming (prosuming) assets connected to the Internet of Things (IoT) to achieve a common goal
through a collaborative effort. This campus-wide test-bed is named Carnegie$\mathbb{PLUG}$, Carnegie Mellon \textbf{P}rosumer-in-the-\textbf{L}oop sim\textbf{U}lation \textbf{G}rid.
Reliable and scalable sensing, networking and actuation  is critical for sustaining operations of porsuming DERs in normal and emergency conditions of the power grid with the added benefit of supporting plug-and-play operation. 
The applications of the proposed test-bed, while illustrated in the context of DER coordination, covers a wide-range of distributed control and machine learning test cases. 

\subsection{Literature Review}
\vspace{-.3cm}

Sensing and actuation test-beds are critical for understanding operational constraints and impacts of small-scale distributed energy resources on the future electric grids. This section reviews capabilities and limitations of existing test-beds that are designed to test strategies for integrating IoT-connected energy resources in power system operations.
A vast body of work have focused on setting up and evaluating multi-agent sensing and control test-beds for energy applications. In this regards, \cite{goyal2016agent} implemented a large-scale test-bed for optimizing energy usage of HVAC (Heating, Ventilating, and Air-Conditioning) and lighting systems across nine buildings. Particularly, this paper tested multiple distributed control coordination methodologies to limit the fan power consumption of air-handling units.
Reference \cite{chinde2016volttron} focused on tackling HVAC control problems on a large-scale multi-zone environment using distributed optimization framework and demonstrated the resulting algorithms on their test bed. 
Authors in \cite{luo2017application} created a small-scale hardware setup to simulate decentralized electricity generation and consumption using sixteen Raspberry Pi single-boarded computers. 
Researchers from Carnegie Mellon university, have developed a campus-wide sensing, networking and actuation platform (named as SensorANDREW) to study generic IoT applications \cite{rowe2011sensor}. Our proposed Carnegie$\mathbb{PLUG}$ test-bed is inspired by SensorANDREW.

\begin{figure*}[t]
    \centering
    \includegraphics[width=5in]{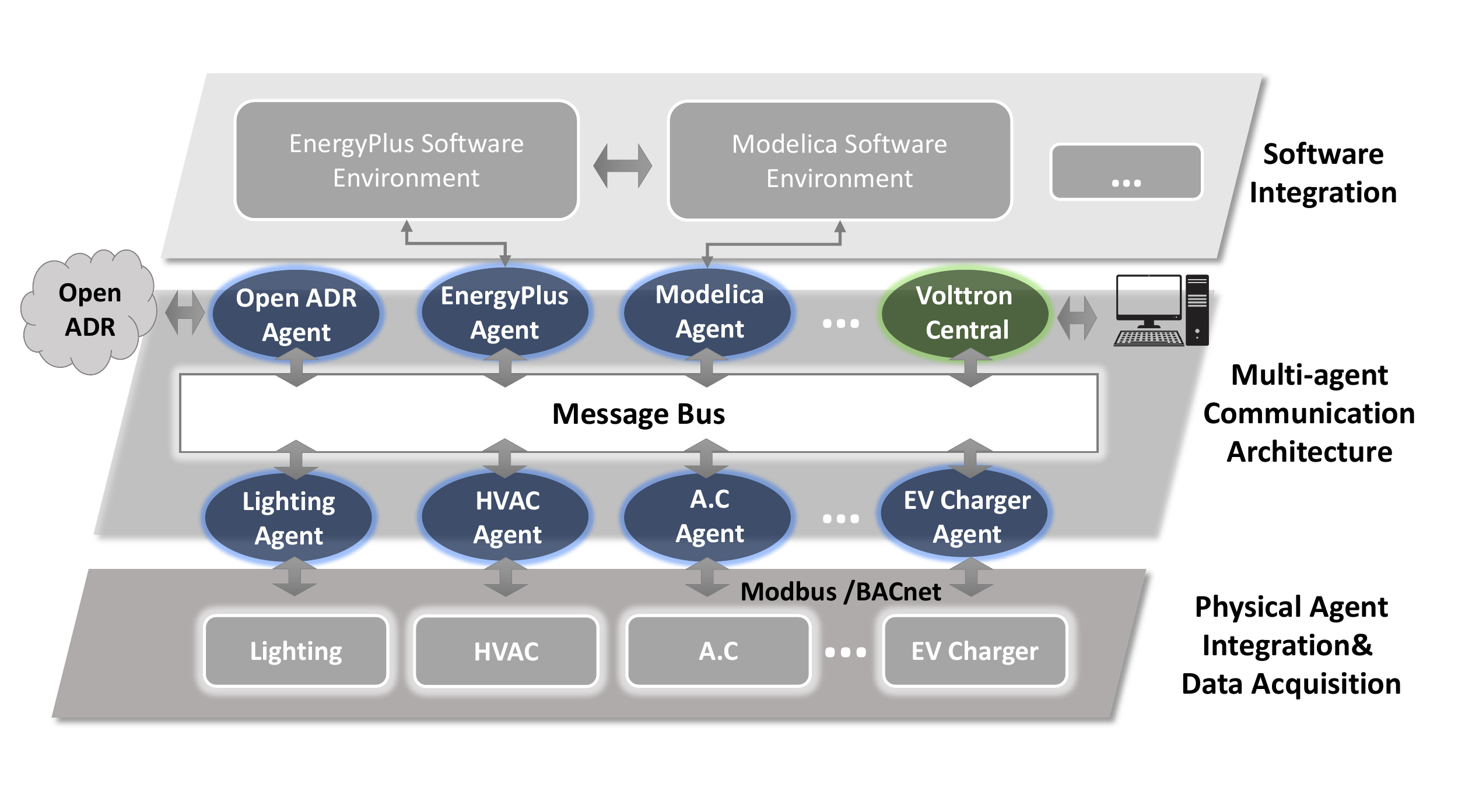}
        \caption{Multi-Layer Architecture of CarnegiePLUG}
    \label{fig:multi-layer-arch}

\end{figure*}


Interoperable and scalable communication platforms are a key enabler of any IoT test-bed.
As heterogeneous devices on test-beds use different protocols, middleware-based solutions are widely adopted to act as an interface between the different devices and protocols such as the open unified connectivity unified architecture (OPC UA) \cite{grossmann2008opc} and OSI Soft \cite{balac2013large}. Numerous energy management platforms have adopted $\text{VOLTTRON}^{\text{TM}}$, an open-source data acquisition and control system \cite{volttronLINK}, to facilitate communication between various energy consuming and producing (prosuming) assets. Authors in \cite{7028784} proposed a $\text{VOLTTRON}^{\text{TM}}$-based conceptual architecture enabling building energy management studies on Virginia Tech campus. Also, \cite{rangel2016integrating} presented details about their Hardware-in-the-Loop test-bed and discussed interactions between the distribution system simulation tool and the $\text{VOLTTRON}^{\text{TM}}$ platform and how this agent-based communication platform facilitates information (weather and power data) exchange between agents/assets. Furthermore, \cite{6799774} have studied using $\text{VOLTTRON}^{\text{TM}}$ for coordinating electric vehicle charging with residential energy use. Moreover, \cite{osti_1326152} reviewed a broad range of applications and deployments and commercialization efforts in small and medium-sized commercial buildings using $\text{VOLTTRON}^{\text{TM}}$. Additionally, \cite{Haack:2013:VAP:2484920.2485228} explored the possibility of using $\text{VOLTTRON}^{\text{TM}}$ for demand response management in a smart grid environment. 

Hardware-based simulations for pursuing energy management objectives are often challenging to setup. On the other hand, software-based simulations are more accessible and faster to develop. 
In this regard, $\text{EnergyPlus}^{\text{TM}}$ \cite{crawley2001energyplus} has been used for many building energy simulation research studies. For example, \cite{doi:10.1080/19401493.2014.891656} presented an $\text{EnergyPlus}^{\text{TM}}$ model-based predictive control for minimizing energy consumption of an HVAC system while maintaining occupant thermal comfort.
Moreover, \cite{wetter2011co} studied the use of the Building Controls Virtual Test Bed (BCVTB) for enabling and connecting different simulation programs for information exchange. 
Finally, \cite{pang2011real} presented a simulation-based  building performance monitoring tool to predict energy consumption.

\subsection{Contributions}
This paper introduces Carnegie$\mathbb{PLUG}$ which stands for \underline{P}rosumer-in-the-\underline{L}oop simu\underline{L}ation \underline{G}rid. 
The Carnegie$\mathbb{PLUG}$ is a campus-wide test-bed designed for multi-agent simulation of IoT-connected prosuming assets. We consider a wide-range of prosuming devices including energy producers (e.g. solar PV) and consumers (e.g. heating and cooling loads). The Carnegie$\mathbb{PLUG}$ ecosystem is comprised of interconnected virtual asset (simulated in software packages) as well as physical devices deployed in different locations on Carnegie Mellon's campus. As presented in Fig. \ref{fig:multi-layer-arch}, Carnegie$\mathbb{PLUG}$ is a multi-layer framework that enables multi-agents large-scale sensing, control, optimization and machine learning simulations.

The data acquisition layer is built on the campus' advanced sensing infrastructure that collects fine-grained environmental (e.g. room temperatures) and energy data (e.g. sub-metering) from most campus buildings and facilities. This system provides real-time data access through Carnegie$\mathbb{PLUG}$'s communication architecture. The test-bed communication architecture is built on the Department of Energy supported open source $\text{VOLTTRON}^{\text{TM}}$ framework. This layer hosts agents and facilitates inter-agent communications. The software integration layer is comprised of interfaces of different software including $\text{EnergyPlus}^{\text{TM}}$ and Modelica. Carnegie$\mathbb{PLUG}$'s multi-layer structure allows modeling of different IoT-connected prosuming devices in different software environments (each as a separate agent) and enables system-wide coordination across numerous agents. This test-bed can be used for a wide-range of research studies including but not limited to distributed multi-agent information processing, federated learning over IoT devices, and transactive energy markets. In what follows, we will describe data acquisition, communication structure and software integration layers of the Carnegie$\mathbb{PLUG}$ framework. Finally, we will present preliminary simulation results to demonstrate the potential of our test-bed.

\section{Data acquisition}
\renewcommand\arraystretch{1}
\begin{table*}[t]
\centering
\caption{Common data points available for real-time monitoring with PI DataLink.}
\begin{tabular}{p{4cm}|p{4cm}|p{3.5cm}|p{4cm}} 
 \toprule
 \textbf{Building Electricity Meters} & \textbf{Building Cooling} & \textbf{Building Heating} & \textbf{Room Level}\\ 
 \midrule
 Apparent Power (VA) & Chilled Water Return and Supply Temperature & Steam Flowrate & Airflow\\ 
 kWh Accumulator & Chilled Water Valves Position & Hot Water Valves Position & Airflow Setpoint\\ 
 Reactive Power (VAR) & Supply Air Static Pressure & Hot Water Supply and Return Temperature & Cooling Setpoint\\ 
 Real Power (kW) & Supply and Return Air Temperature & & Damper Position Command\\
  & AHU Run Status & & Discharge Temperature \\ 
  & & & Heating Setpoint \\ 
  & & & Hot Water Valve Command \\ 
  & & & Occupancy Status \\ 
  & & & Space Temperature \\ 
 \bottomrule 
\end{tabular}
\label{table:1}
\end{table*}

Data acquisition is the underlying step for performing hardware-in-the-loop data analysis of prosuming assets.
In most Carnegie Mellon buildings, the energy and environmental data, such as electricity consumption and temperature, are collected through the campus' advanced sensing system. The primary data collection platform on CMU's campus is the PI-Server program. This program collects real-time building information using a variety of communication protocols and uses PI Vision as the visualization tool. In addition, we are leveraging the legacy sensing infrastructure of Carnegie Mellon's Sensor ANDREW \cite{rowe2011sensor} (e.g., monitoring zone occupancy). The data collected from the these campus sensors cannot be directly utilized for optimized control of lighting and HVAC systems. To this end, we developed a data acquisition layer to connect data collected from the sensing system to the communication channels. This will eventually enable us to efficiently control the energy prosuming devices on campus.
The integration of sensing systems and Carnegie$\mathbb{PLUG}$ will be done through connection of  PI DataLink and $\text{VOLTTRON}^{\text{TM}}$.


\begin{figure}[t]
    \centering
    \includegraphics[width=3.3in]{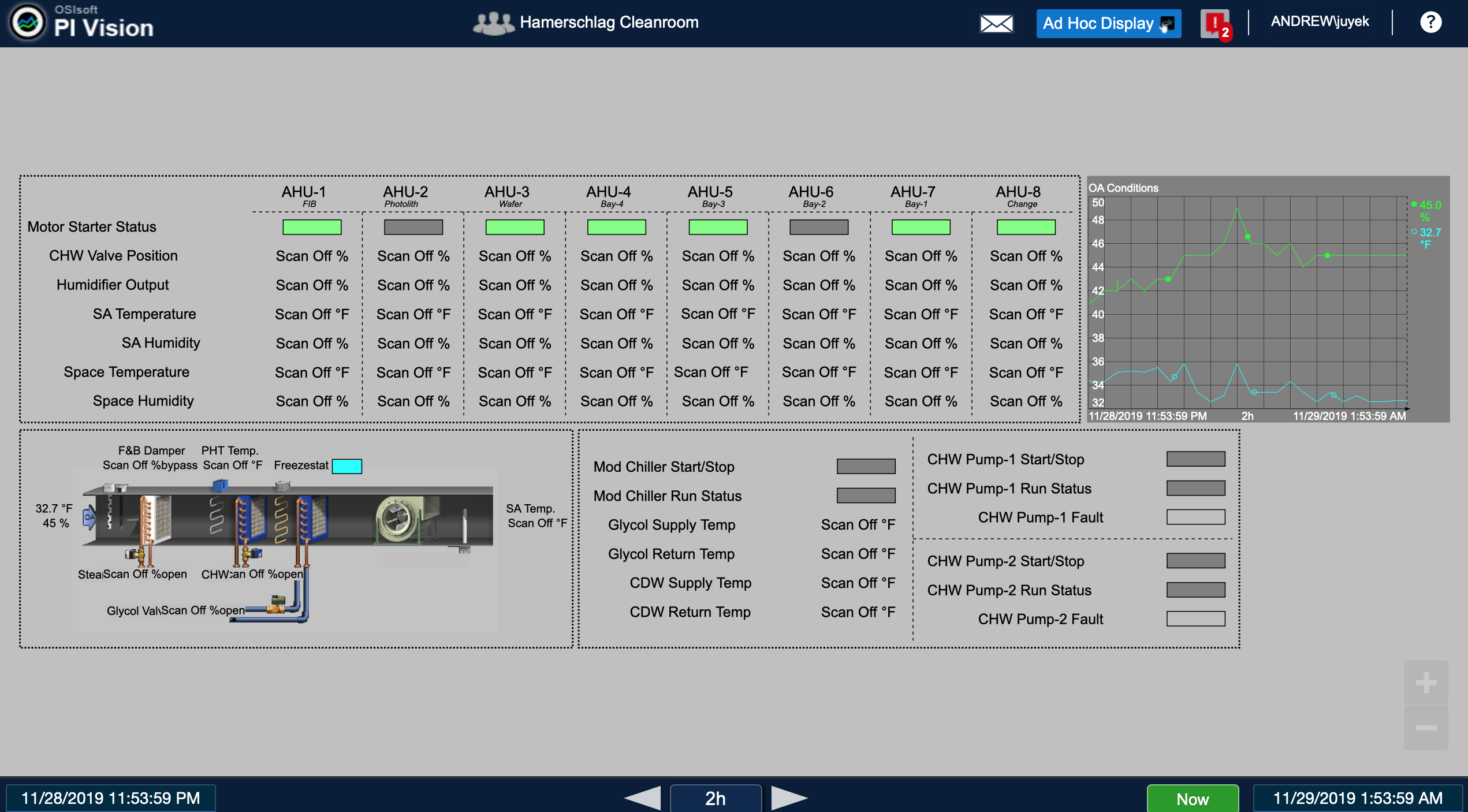}
        \caption{PI Vision of Carnegie Mellon University Campus}
    \label{fig:PI-Coresight}

\end{figure}

PI Vision, shown in Fig. \ref{fig:PI-Coresight}, is a web-based data visualization system that presents real-time and historical operational data collected from various data sources including SCADA and DCS equipment. Carnegie Mellon University's Facility Management Services offers complete access to the SCADA system's datapoints via PI Server and visualization and data analysis with PI Vision and the Excel plug-in PI DataLink. PI Server aggregates sensor information to a singular platform using BACnet, Modbus, HTML web-scraping, relational database harvesting, and CSV file parsing interfaces with BACnet being the most common protocol. The PI DataLink Excel plug-in allows the user to browse data from various buildings and campus wide systems (e.g. campus steam system and snowmelt system) and select data points to monitor. Historical data is selected by specifying the time range and time interval for the PI data points chosen. Real-time data can be monitored and updated at a specified rate. Historical data collected through the Excel PI DataLink plug-in can be exported to software environments. For example, this data can be exported to 
the $\text{EnergyPlus}^{\text{TM}}$ file (IDF) for building calibration to obtain more accurate simulation results of campus buildings. Real-time data collected will be autosaved to a CSV file and read by the $\text{VOLTTRON}^{\text{TM}}$ communication architecture. 

The sensing data available vary between buildings due to different system ages. Common data points for most campus building are outlined in Table \ref{table:1}. Specifically, Porter Hall's new B-Level office space provides more comprehensive HVAC data for each room. Real-time occupancy and thermostat data can be read by the $\text{VOLTTRON}^{\text{TM}}$ agent and exported to $\text{EnergyPlus}^{\text{TM}}$ or Modelica for a coupling of a simulated building environment with real-time data. This allows for testing control algorithms without impacting the physical building's operations.



$\text{VOLTTRON}^{\text{TM}}$ provides direct access to heterogeneous devices and assets (e.g., HVAC units, power systems, and electronic distribution systems) through standard protocols such as BACNet and Modbus. Each device has its own driver agent within a $\text{VOLTTRON}^{\text{TM}}$ instance. The driver agent is responsible for interfacing with it's associated device. Each driver agent is created and executed based on the master driver agent in $\text{VOLTTRON}^{\text{TM}}$. Each driver agent has a configuration file that stores information about the device such as IP address, port number, and communication interface type (e.g., Modbus). With this information, each driver agent facilitates real time data exchange with the physical device it's responsible for.
The basic mechanism behind data collection for each driver agent is a heartbeat-like signaling process. The heartbeat-like signaling is a recurrent messaging that provides an entry to send and receive periodic messages from a driver agent to its associated device. Set to send data request every $n$ seconds, the driver agent sends and receives a data packet from the campus sensors and makes the packet available to other components of Carnegie$\mathbb{PLUG}$ through the message bus, or the Information Exchange Bus (IEB). The next section presents more details regarding information exchange through the message bus.

\section{communication architecture}
\label{communication-architecture}

The core of Carnegie$\mathbb{PLUG}$ is built on a multi-agent communication layer that enables Carnegie$\mathbb{PLUG}$ components to send and receive messages (i.e. data, topic and control signals) in real-time. In this communication layer, agents represent individual components of Carnegie$\mathbb{PLUG}$. The agent is essentially an autonomous program that reads and writes messages to the message bus and performs procedures based on the preset conditions or messages it receives. Each asset that is plugged into Carnegie$\mathbb{PLUG}$ interacts with other components through their agents residing in the communication layer. All agents communicate through the message bus. 

Message exchange follows particular protocols. Each message has topics and each agent subscribes to certain topics. When an agent publishes (send) a message to the bus, it should denote a topic for that message. Upon publishing new messages, those agents that have subscribed to that message's topic will pull the message and triggers a callback function defined on the topic. 

The foundation of our test-bed's communication architecture is based on the $\text{VOLTTRON}^{\text{TM}}$ platform, which is a distributed control and sensing platform that runs on the top of the TCP/IP-based internet communication protocol. The multi-agent communication structure allows us to connect different buildings using the existing LAN/Wifi network 
and develop large-scale multi-building test-bed. Furthermore, by supporting standard industrial electronic communication protocols such as BACnet and Modbus, we are able to host heterogeneous energy prosuming assets ranging from HVAC equipment to solar energy generation systems on the campus. The chosen communication setup is open sources and offers transparency, scalability and customization.

The Carnegie$\mathbb{PLUG}$ test-bed has the capability of adjusting electric power consumption in response to fluctuations in grid demand through OpenADR (Automated Demand Response). The OpenADR is a Department of Energy supported standard specification that enables electricity providers to send information about the current price of electricity and reliability of the grid directly to customers over the Internet. Hence, electricity consumers like the campus can manage and control the consumption of the current energy usage regarding price changes. The Carnegie$\mathbb{PLUG}$ and OpenADR communicate through their agents in $\text{VOLTTRON}^{\text{TM}}$ environment.

\vspace{-.2cm}
\section{Modeling}

 We use a wide range of software environments to model operational constraints of different energy prosumers and derive optimal actuation set-points for these assets. The software packages will interface with their associated agent in $\text{VOLTTRON}^{\text{TM}}$. The reminder of this section will provide integration details of two widely used building load modeling software with our test-bed.


\vspace{-.1cm}
\subsection{EnergyPlus}

$\text{EnergyPlus}^{\text{TM}}$, from the Department of Energy's Building Technology Office, is an open-source software that offers full building energy simulations. Simulation results include whole building energy consumption and load, HVAC zonal variables, equipment energy consumption and load, and various building performance metrics. At a high-level, $\text{EnergyPlus}^{\text{TM}}$ input files require a building envelope to define the building, a weather file, and an HVAC system \cite{getting_started}. The building's zones and loads are defined in the Input Data File (IDF) through objects (Schedule object, Lights object, Refrigeration Object, etc.). $\text{EnergyPlus}^{\text{TM}}$ developers provide over 700 Example Files for various applications and can be modified to fit the user's specifications. 

Simulating a Carnegie Mellon building requires definition of the building envelope and equipment. 2-D construction drawings can be imported into the DesignBuilder software and modified to specify HVAC zones. DesignBuilder allows for easy integration with the $\text{EnergyPlus}^{\text{TM}}$ IDF where building loads are defined for each HVAC zone identified in the DesignBuilder model. \cite{praprost_investigating_2018}

$\text{EnergyPlus}^{\text{TM}}$ offers co-simulation and external control through the Building Control Virtual Test Bed (BCVTB). The External Interface object within the $\text{EnergyPlus}^{\text{TM}}$ IDF allows external control of schedules, variables, and actuators (control status of components). The simulation results are read at each simulation time-step by the external device through definition of the Output Variables object. External control allows for more sophisticated control algorithms than the internal Energy Management System (EMS) object and equipment schedules can be altered to reduce the peak demand of interconnected systems that a standalone $\text{EnergyPlus}^{\text{TM}}$ simulation cannot account for. \cite{external_int}

A $\text{VOLTTRON}^{\text{TM}}$ $\text{EnergyPlus}^{\text{TM}}$ agent is created to facilitate communication and apply external control to the $\text{EnergyPlus}^{\text{TM}}$ simulations that are dependent on other system loads and operations. This agent communicates with $\text{EnergyPlus}^{\text{TM}}$ over a local BSD socket connection. The variables defined in the Output Variables object are mapped to PubSub topics in the EnergyPlus Agent configuration file. Other agents are able to subscribe to the topics the EnergyPlus Agent publishes data to, and read the data at each simulation time-step. A Control Agent determines the appropriate action for the parameters defined in the External Interface object and publishes the controls to a topic that the EnergyPlus Agent subscribes to. The control inputs are mapped to the External Interface object names within the EnergyPlus Agent configuration file and the $\text{EnergyPlus}^{\text{TM}}$ simulation runs with the altered parameters. \cite{corbin_co-simulation_2017} 

Including $\text{EnergyPlus}^{\text{TM}}$ simulations in Carnegie$\mathbb{PLUG}$ allows for more comprehensive testing of controls as well as a more diverse set of assets the control system can incorporate into its decision process. $\text{EnergyPlus}^{\text{TM}}$ models can be as simple as a single room with a lighting load and a refrigerator, to as large scale as a Carnegie Mellon campus building. Through a $\text{VOLTTRON}^{\text{TM}}$ CSV Reader Agent for PI Datalink, the EnergyPlus Agent can subscribe to the topic the CSV Reader Agent publishes data to. An $\text{EnergyPlus}^{\text{TM}}$ simulation file for a Carnegie Mellon campus building can react in real-time to collected data (e.g., occupancy and thermostat settings), as well as equipment schedules.

\subsection{Modelica}
    \vspace{-.05cm}

Modelica with Modelica's {\fontfamily{qcr}\selectfont
Buildings} Library is another simulation engine for building energy modeling. Modelica has more robust control capabilities and can be coupled with $\text{EnergyPlus}^{\text{TM}}$ simulations as well as run as a standalone application. The Package {\fontfamily{qcr}\selectfont
Controls} within the {\fontfamily{qcr}\selectfont
Buildings} Library has predefined control blocks to be used in the building energy simulation. Modelica uses the BCVTB as the software link to connect the engine with external applications. Co-simulation with $\text{EnergyPlus}^{\text{TM}}$ requires the import and export of Functional Mockup Units from Modelica and $\text{EnergyPlus}^{\text{TM}}$ to the BCVTB to share model information and allow for data exhange and control. \cite{Wetter} 

The $\text{VOLTTRON}^{\text{TM}}$-based communication channel of Carnegie$\mathbb{PLUG}$ exchanges infromation with Modelica through the Python Socket package which facilitates communication between two Python programs over TCP/IP protocol. Modelica's {\fontfamily{qcr}\selectfont
Buildings} Library includes a Python interface which is how the program connects to the Python Socket. Similar to the EnergyPlus Agent, the Modelica Agent publishes and subscribes to topics in the $\text{VOLTTRON}^{\text{TM}}$ message bus and modifies parameters in Modelica based on the information received from the $\text{VOLTTRON}^{\text{TM}}$ message bus. \cite{HIL_2018}

    \vspace{-.05cm}
\section{Physical interfaces and agent definitions}
    \vspace{-.05cm}
In addition to interfacing with virtual agents in simulation environments, our test-bed seamlessly connects with controller of physical assets. Our current setup relies on Programmable Logic Controllers (PLCs) and Raspberry Pis.


PLCs are rugged, industrial controlers that are used in a wide variety of systems in need of controls. Inclusion of PLC hardware in Carnegie$\mathbb{PLUG}$, ensures that our communication architecture can be implemented in more general applications.
The PLC program can emulate a vehicle charging station, a hot water heater, or any energy prosuming equipment with a known energy profile. PLCs can also simulate a home environment with a Human Machine Interface (HMI) for localized control of the home system. The user can locally (at the HMI) plug-in an electric vehicle and watch how the $\text{VOLTTRON}^{\text{TM}}$ control architecture reacts to the additional load on the system. For example, by prioritizing different loads, the water heater may be shut-off if there is limited impact to the homeowner's service  \cite{6799774}. PLCs also replace physical relays and can therefore control physical assets through relay logic in addition to analog outputs. This enables $\text{VOLTTRON}^{\text{TM}}$ to control assets without networked communication architecture in place. Data collection of physical assets through PLCs include analog inputs (4-20mA signals or 0-10V signals) and digital inputs from a physical asset's relay outputs.

\begin{figure}[t]
    \centering
    \includegraphics[width=3.4in]{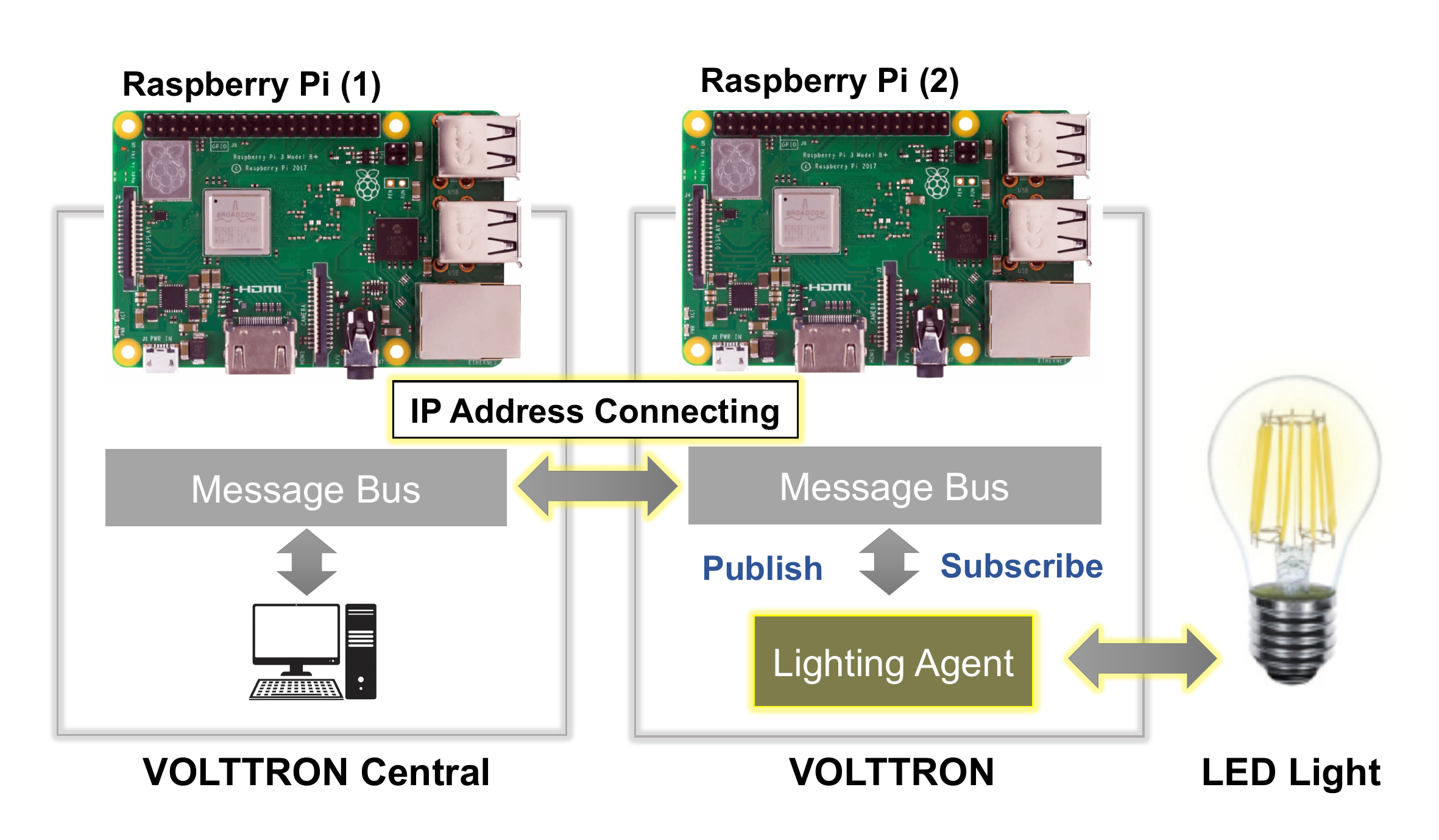}
        \caption{Information Exchange Between Two Raspberry Pis}

    \label{fig:RaspberryPi-Volttron-2}
    \vspace{-.5cm}
\end{figure}

In addition to PLCs, the well-adopted Raspberry Pi controller is included in Carnegie$\mathbb{PLUG}$. Raspberry Pi is a low cost, credit-card sized computing hardware platform that consists of CPU, main board, memory, input-output ports and network interfaces. Networked Raspberry Pis are key enabler of Carnegie$\mathbb{PLUG}$'s sensing, local computation and actuation functionalities. 
The roles of Raspberry Pi in our current setup include (i) hosting $\text{VOLTTRON}^{\text{TM}}$ Central agent that can monitor and control any number of agents (ii) enabling on-site computations for energy prosuming devices (iii) collecting data and connecting physical and virtual environments 
The data  collected from the physical devices can be transmitted to the virtual environment through the device-specific agent running on the $\text{VOLTTRON}^{\text{TM}}$ instance in the Raspberry Pi environment.  Also, control signals from virtual environments can be delivered to physical devices through the agents on Raspberry Pi. Figure \ref{fig:RaspberryPi-Volttron-2} illustrates information exchange between two Raspberry Pis. 

    \vspace{-.1cm}
\section{Preliminary results}



\begin{figure}[t]
    \centering
    \includegraphics[width=3.3in]{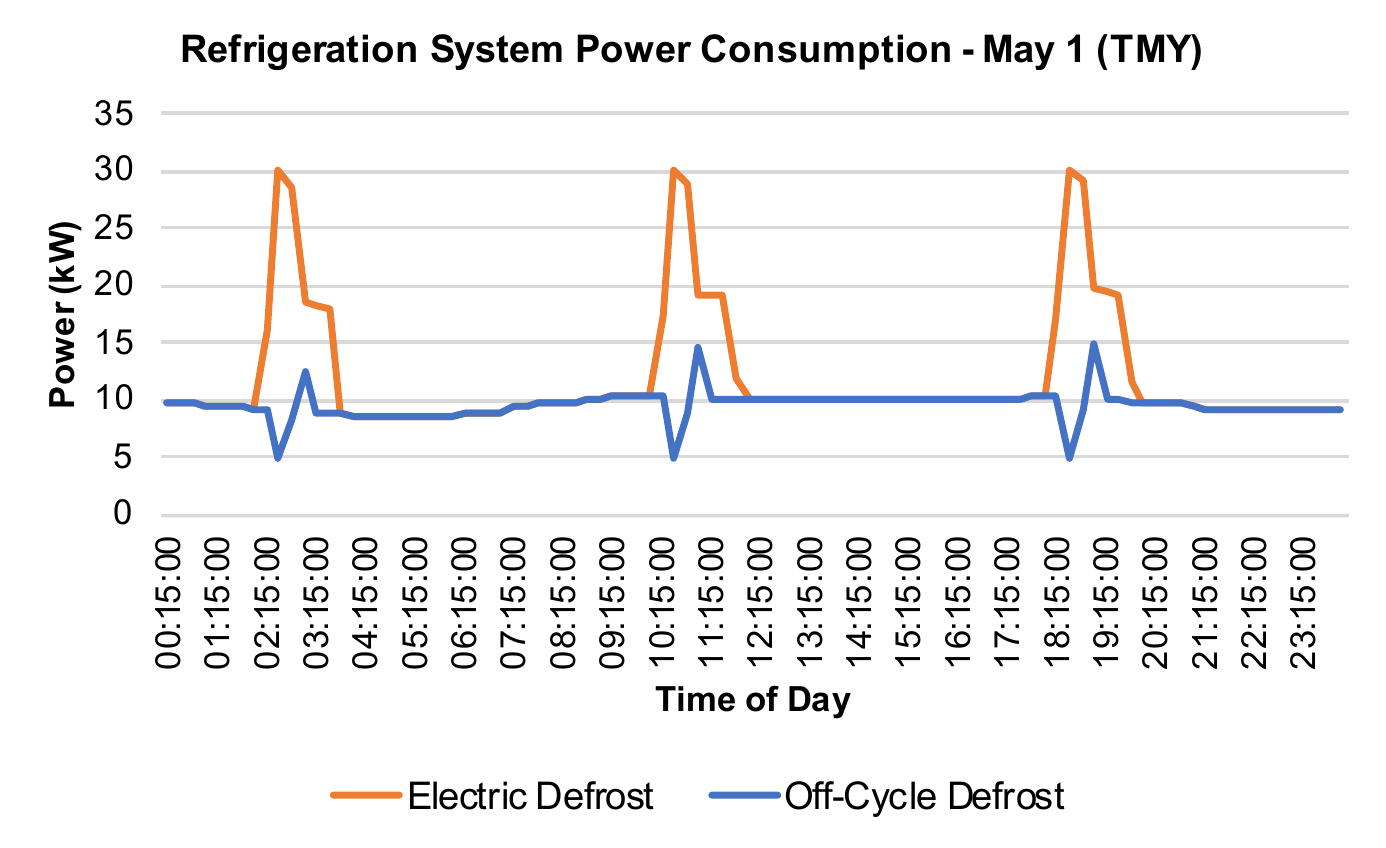}
    \caption{$\text{EnergyPlus}^{\text{TM}}$ Three Defrost Cycles Power Consumption}
    \label{fig:Three Defrost}
        \vspace{-.5cm}
\end{figure}

This section is devoted to showcasing preliminary energy simulation results and the potential of using Carnegie$\mathbb{PLUG}$ to control energy prosuming assets.
We used $\text{EnergyPlus}^{\text{TM}}$ software to schedule defrost of a commercial walk-in refrigerator illustrating the importance of load monitoring and peak shaving. The focus of the simulation is on how the walk-in refrigerator's power consumption responds to the on/off schedule of the defrost system. Therefore, the simulation was a single room with the only system load being the walk-in refrigerator. Figure. \ref{fig:Three Defrost} shows the power consumption for the walk-in refrigeration system for May 1 (using Typical Meteorological Year weather data) with electric defrost and off-cycle defrost (i.e. the system is turned off to allow for defrost) which were scheduled from 2:12-2:42am, 10:12-10:42am, and 6:12-6:42pm. The defrost schedule was taken from the actual defrost times of a refrigerator unit at the Cohon University Center at Carnegie Mellon. The electric defrost increases power consumption during the defrost cycle and directly after when the compressor needs to compensate for the added heat to the system. The off-cycle defrost system reduces power consumption during defrost but increases power consumption directly after due to the compressor running. 

\begin{figure}[t]
    \centering
    \includegraphics[width=3.3in]{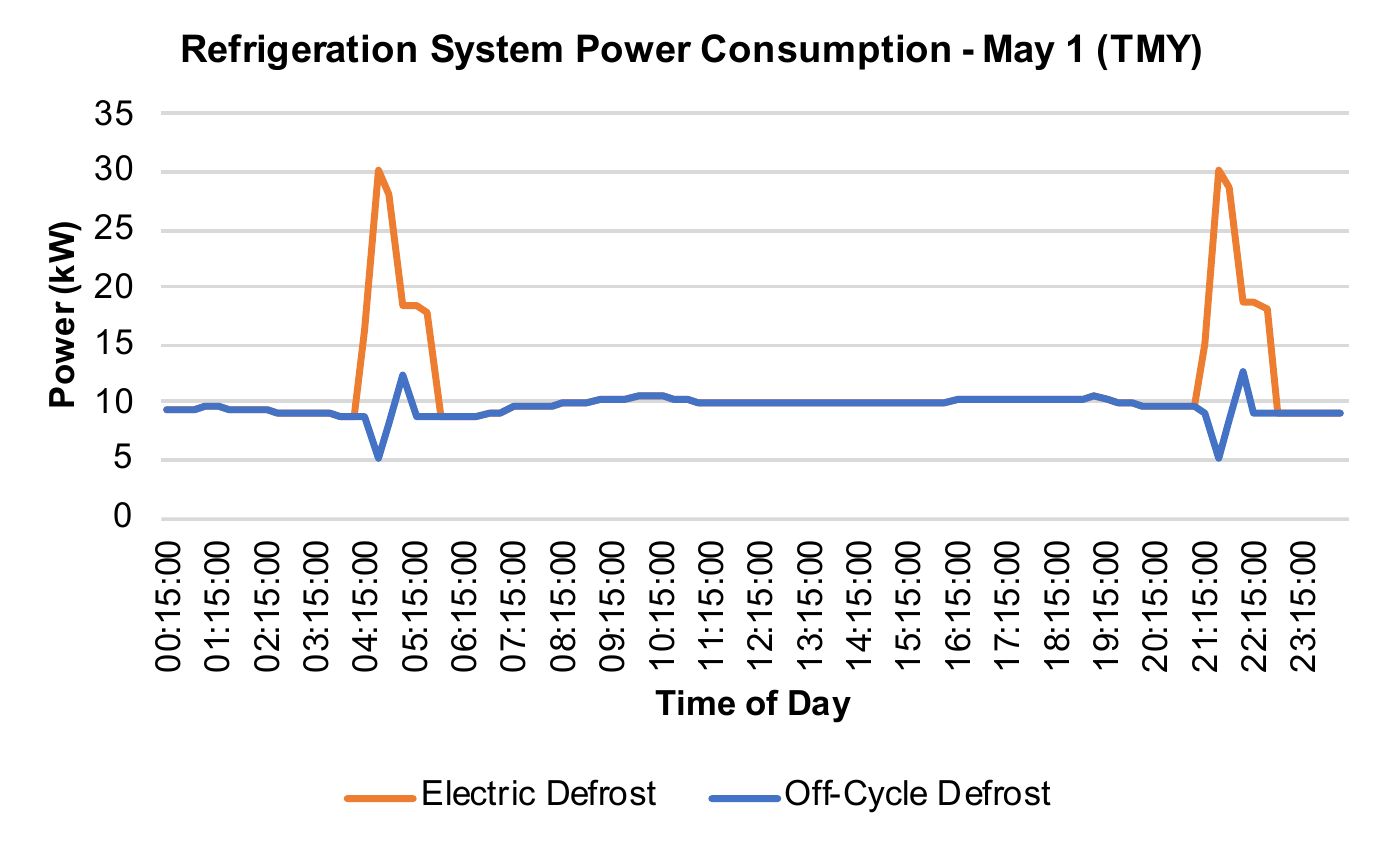}
        \caption{$\text{EnergyPlus}^{\text{TM}}$ Two Shifted Defrost Cycles Power Consumption}
    \label{fig:Two Defrost}
    
    \vspace{-.6cm}
\end{figure}

Scheduling electric defrost during periods of peak demand is problematic for supermarkets or buildings with large refrigeration loads. The addition of the electric defrost power consumption to the peak demand of the building increases their monthly demand charge. By shifting the defrost load to periods of overall low power consumption, the building owner or tenant can save money and also lessen stress on the grid. Figure. \ref{fig:Two Defrost} shifts the defrost cycles to the early morning (4:12-4:42am) and late evening (9:12-9:42pm) to illustrate how this would be done. For multiple refrigeration units with off-cycle defrost systems, the defrost of each refrigerator could be scheduled so they negate one another. For example, while one refrigeration unit is in defrost, another unit will have just finished defrost and have a spike in power consumption. The reduction in power consumption from one unit can cancel out the increase in power consumption of the other unit in post-defrost. Coupling the $\text{EnergyPlus}^{\text{TM}}$ refrigeration simulations with $\text{VOLTTRON}^{\text{TM}}$ allows for further coordination of defrost schedules with other physical and virtual assets for the desired power reduction. 

    \vspace{-.4cm}
\section{Conclusion}
    \vspace{-.1cm}
In this paper, we presented a campus-wide multi-agent sensing and decision-making framework that integrates a wide range of IoT-connected DERs. The plug-and-play architecture of this multi-agent test-bed called, Carnegie$\mathbb{PLUG}$, facilitates integration of heterogeneous energy prosuming assets. The test-bed interfaces with different DER modeling simulators and communication protocols. We intend to extend sensing and control capabilities of this test-bed and leverage it to pursue a variety of research objectives.


\bibliography{bibfile}{}

\begin{thebibliography}{10}
\providecommand{\url}[1]{#1}
\csname url@samestyle\endcsname
\providecommand{\newblock}{\relax}
\providecommand{\bibinfo}[2]{#2}
\providecommand{\BIBentrySTDinterwordspacing}{\spaceskip=0pt\relax}
\providecommand{\BIBentryALTinterwordstretchfactor}{4}
\providecommand{\BIBentryALTinterwordspacing}{\spaceskip=\fontdimen2\font plus
\BIBentryALTinterwordstretchfactor\fontdimen3\font minus
  \fontdimen4\font\relax}
\providecommand{\BIBforeignlanguage}[2]{{%
\expandafter\ifx\csname l@#1\endcsname\relax
\typeout{** WARNING: IEEEtran.bst: No hyphenation pattern has been}%
\typeout{** loaded for the language `#1'. Using the pattern for}%
\typeout{** the default language instead.}%
\else
\language=\csname l@#1\endcsname
\fi
#2}}
\providecommand{\BIBdecl}{\relax}
\BIBdecl

\bibitem{goyal2016agent}
S.~Goyal, W.~Wang, and M.~R. Brambley, ``An agent-based test bed for building
  controls,'' in \emph{2016 American Control Conference (ACC)}.\hskip 1em plus
  0.5em minus 0.4em\relax IEEE, 2016, pp. 1464--1471.

\bibitem{chinde2016volttron}
V.~Chinde, A.~Kohl, Z.~Jiang, A.~Kelkar, and S.~Sarkar, ``A volttron based
  implementation of supervisory control using generalized gossip for building
  energy systems,'' 2016.

\bibitem{luo2017application}
J.~Luo, H.~Pourbabak, and W.~Su, ``The application of distributed control
  algorithms using volttron-based software platform,'' in \emph{2017 8th
  International Renewable Energy Congress (IREC)}.\hskip 1em plus 0.5em minus
  0.4em\relax IEEE, 2017, pp. 1--6.

\bibitem{rowe2011sensor}
A.~Rowe, M.~E. Berges, G.~Bhatia, E.~Goldman, R.~Rajkumar, J.~H. Garrett, J.~M.
  Moura, and L.~Soibelman, ``Sensor andrew: Large-scale campus-wide sensing and
  actuation,'' \emph{IBM Journal of Research and Development}, vol.~55, no.
  1.2, pp. 6--1, 2011.

\bibitem{grossmann2008opc}
D.~Grossmann, K.~Bender, and B.~Danzer, ``Opc ua based field device
  integration,'' in \emph{2008 SICE annual conference}.\hskip 1em plus 0.5em
  minus 0.4em\relax IEEE, 2008, pp. 933--938.

\bibitem{balac2013large}
N.~Balac, T.~Sipes, N.~Wolter, K.~Nunes, B.~Sinkovits, and H.~Karimabadi,
  ``Large scale predictive analytics for real-time energy management,'' in
  \emph{2013 IEEE international conference on big data}.\hskip 1em plus 0.5em
  minus 0.4em\relax IEEE, 2013, pp. 657--664.

\bibitem{volttronLINK}
``Volttron documentation,'' \emph{available at: https://volttron.org/}.

\bibitem{7028784}
W.~{Khamphanchai}, A.~{Saha}, K.~{Rathinavel}, M.~{Kuzlu},
  M.~{Pipattanasomporn}, S.~{Rahman}, B.~{Akyol}, and J.~{Haack}, ``Conceptual
  architecture of building energy management open source software (bemoss),''
  in \emph{IEEE PES Innovative Smart Grid Technologies, Europe}, Oct 2014, pp.
  1--6.

\bibitem{rangel2016integrating}
R.~Rangel, ``Integrating volttron and gridlab-d to create a power
  hardware-in-the-loop test bed,'' 2016.

\bibitem{6799774}
J.~{Haack}, B.~{Akyol}, N.~{Tenney}, B.~{Carpenter}, R.~{Pratt}, and
  T.~{Carroll}, ``Volttron™: An agent platform for integrating electric
  vehicles and smart grid,'' in \emph{2013 International Conference on
  Connected Vehicles and Expo (ICCVE)}, Dec 2013, pp. 81--86.

\bibitem{osti_1326152}
K.~A. Cort, J.~N. Haack, S.~Katipamula, and A.~K. Nicholls, ``Volttron™:
  Tech-to-market best-practices guide for small- and medium-sized commercial
  buildings,'' 7 2016.

\bibitem{Haack:2013:VAP:2484920.2485228}
\BIBentryALTinterwordspacing
J.~Haack, B.~Akyol, B.~Carpenter, C.~Tews, and L.~Foglesong, ``Volttron: An
  agent platform for the smart grid,'' in \emph{Proceedings of the 2013
  International Conference on Autonomous Agents and Multi-agent Systems}, ser.
  AAMAS '13.\hskip 1em plus 0.5em minus 0.4em\relax Richland, SC: International
  Foundation for Autonomous Agents and Multiagent Systems, 2013, pp.
  1367--1368. [Online]. Available:
  \url{http://dl.acm.org/citation.cfm?id=2484920.2485228}
\BIBentrySTDinterwordspacing

\bibitem{crawley2001energyplus}
D.~B. Crawley, L.~K. Lawrie, F.~C. Winkelmann, W.~F. Buhl, Y.~J. Huang, C.~O.
  Pedersen, R.~K. Strand, R.~J. Liesen, D.~E. Fisher, M.~J. Witte
  \emph{et~al.}, ``Energyplus: creating a new-generation building energy
  simulation program,'' \emph{Energy and buildings}, vol.~33, no.~4, pp.
  319--331, 2001.

\bibitem{doi:10.1080/19401493.2014.891656}
\BIBentryALTinterwordspacing
J.~Zhao, K.~P. Lam, B.~E. Ydstie, and O.~T. Karaguzel, ``Energyplus model-based
  predictive control within design–build–operate energy information
  modelling infrastructure,'' \emph{Journal of Building Performance
  Simulation}, vol.~8, no.~3, pp. 121--134, 2015. [Online]. Available:
  \url{https://doi.org/10.1080/19401493.2014.891656}
\BIBentrySTDinterwordspacing

\bibitem{wetter2011co}
M.~Wetter, ``Co-simulation of building energy and control systems with the
  building controls virtual test bed,'' \emph{Journal of Building Performance
  Simulation}, vol.~4, no.~3, pp. 185--203, 2011.

\bibitem{pang2011real}
X.~Pang, P.~Bhattacharya, Z.~O’Neill, P.~Haves, M.~Wetter, and T.~Bailey,
  ``Real-time building energy simulation using energyplus and the building
  controls virtual test bed,'' in \emph{Proceeding of the 12th IBPSA
  Conference}, 2011, pp. 2890--2896.

\bibitem{getting_started}
``Getting {Started},'' \emph{EnergyPlus Version 9.2.0}, pp. 34--35, Sept. 2019.

\bibitem{praprost_investigating_2018}
M.~A. Praprost, ``\BIBforeignlanguage{en}{{Investigating} {EnergyPlus} {as} {a}
  {Simulation} {Tool} {for} {Deploying} {VOLTTRON} {Transactive} {Energy}
  {Technologies} {in} {Commercial} {Buildings}},'' p. 135, 2018.

\bibitem{external_int}
``External {Interface}(s) {Application} {Guide},'' \emph{EnergyPlus Version
  9.2.0}, pp. 5--14, Sept. 2019.

\bibitem{corbin_co-simulation_2017}
\BIBentryALTinterwordspacing
C.~D. Corbin, D.~Vrabie, and S.~Katipamula,
  ``\BIBforeignlanguage{en}{Co-{Simulation} and {Validation} of {Advanced}
  {Building} {Controls} with {VOLTTRONTM} and {EnergyPlusTM}},'' pp. 3--5,
  2017. [Online]. Available: \url{https://doi.org/10.26868/25222708.2017.691}
\BIBentrySTDinterwordspacing

\bibitem{Wetter}
\BIBentryALTinterwordspacing
M.~Wetter, W.~Zuo, T.~S. Nouidui, and X.~Pang, ``Modelica buildings library,''
  \emph{Journal of Building Performance Simulation}, vol.~7, no.~4, pp.
  253--270, 2014. [Online]. Available:
  \url{https://doi.org/10.1080/19401493.2013.765506}
\BIBentrySTDinterwordspacing

\bibitem{HIL_2018}
\BIBentryALTinterwordspacing
S.~Huang, W.~Wang, M.~R. Brambley, S.~Goyal, and W.~Zuo, ``An agent-based
  hardware-in-the-loop simulation framework for building controls,''
  \emph{Energy and Buildings}, vol. 181, pp. 26 -- 37, 2018. [Online].
  Available:
  \url{http://www.sciencedirect.com/science/article/pii/S037877881831764X}
\BIBentrySTDinterwordspacing

\end{thebibliography}
\bibliographystyle{IEEEtran}
\end{document}